\documentclass[fleqn,usenatbib]{mnras}

\usepackage[T1]{fontenc}
\usepackage{ae,aecompl}

\usepackage{hyperref}
\usepackage{booktabs}
\usepackage{multirow}
\usepackage{pgf}
\usepackage{subcaption}
\usepackage{threeparttable}
\RequirePackage{fix-cm}

\usepackage{graphicx}	
\usepackage{amsmath}	
\usepackage{amssymb}	

\usepackage{newtxtext,newtxmath}

\title[Excess of edge-on quiescent late-type galaxies]{Too many quiescent late-type galaxies or a matter of misclassification?}

\author[J.\ L.\ Tous et al.]{
J.\ L.\ Tous$^{1}$\thanks{E-mail: j.l.tous-mayol@soton.ac.uk}, S. I. Raimundo$^{1}$ and H. Domínguez Sánchez$^{2}$\\
$^{1}$School of Physics and Astronomy, University of Southampton, Highfield, Southampton SO17 1BJ, UK\\
$^{2}$Instituto de Física de Cantabria, Av. de los Castros, 39005 Santander, Cantabria, Spain\\
}
\date{Accepted XXX. Received YYY; in original form ZZZ}

\pubyear{2025}

\begin{document}
\label{firstpage}
\pagerange{\pageref{firstpage}--\pageref{lastpage}}
\maketitle

\begin{abstract}
In the local Universe, the terms late-type and early-type are used to describe, respectively, disc galaxies with spiral arms and more bulge-dominated systems lacking spiral structure. These morphologies are often associated with different star formation levels, motivating the classical scenario where late-type and early-type galaxies are seen, respectively, as star-forming and “red and dead” systems. To build a comprehensive picture of galaxy evolution, we must understand the relationship between morphology and star formation, and be able to explain more exotic configurations, such as quiescence in late-type galaxies. Large morphological catalogues are valuable tools to address these questions, but they must not be used blindly. We show that selecting quiescent galaxies classified as late‑type in these catalogues leads to a clear excess of edge‑on systems, with unreliable morphologies, compared with the general galaxy population in the Sloan Digital Sky Survey. Consequently, the fraction of quiescent late-type galaxies in this survey could be overestimated by a factor of two. Conversely, S0 galaxies, considered early-type discs without spiral arms, are under-represented at high inclinations. We conclude that the excess of edge-on quiescent late-type galaxies can be explained if most of these systems are misclassified edge-on S0 galaxies. We make recommendations to remove this bias.
\end{abstract}


\section{Introduction}
\label{S:intro}

In the local Universe, galaxies can be classified based on their morphology into late- and early-types \citep{Hubble1926}. In late-type galaxies (LTGs), the main structural component is a rotationally-supported thin disc with conspicuous spiral arms that surrounds a central bulge. In early-type galaxies (ETGs), the contribution from random stellar motions is more important, favouring pressure-supported components where spiral arms cannot easily form. The former type of galaxies tend to be characterised by high star formation levels, whereas the latter are preferentially quiescent systems \citep[e.g.,][]{Strateva2001, Bell2004}. Lenticular (S0) galaxies occupy an intermediate position in this scheme: like LTGs, they host two components---a prominent bulge and a smooth disc---but are generally classified as ETGs because they lack spiral arms and show little or no ongoing star formation.

The existence of this relationship between morphology and star formation activity suggests that the same processes that quench activity in galaxies could also drive a morphological transformation, and vice versa. On one hand, without cold gas to dissipate energy through star formation following the Kennicutt–Schmidt relation \citep{KSlaw1998}, disc galaxies are more likely to evolve toward dynamically hotter, pressure-supported systems with no spiral arms \citep{Sellwood2022}. On the other hand, the growth of a central bulge stabilizes the system against gas fragmentation, hence preventing star formation \citep{Martig2009}. Observations such as the well known morphology-density relation \citep{Dressler1980} along with the increase in the fraction of quiescent galaxies with increasing environmental density \citep[e.g.,][]{Tous2020, JimP22, Lopes2024, Cleland2025, Gort2025} support the view that morphology and galaxy quenching are related \citep[see also][]{GonzalezDelgado2016, Cano-Diaz2019}. In addition, there is evidence suggesting that this relationship is already in place in the high-redshift Universe \citep[e.g.,][]{Wuyts2011}.

However, the link between morphology and star formation could be more complex. For example, there is growing evidence showing that star formation in ETGs is not as rare as previously believed \citep{Kaviraj2007, Salim2012, Mendez-Abreu2019, Tous2020, JimP22, Paspaliaris2023, Tous2025}. In this case, accretion events, such as minor mergers, are thought to be the primary triggers of activity and rejuvenation in ETGs \citep[e.g.,][submitted]{mckelvie2018, Deeley2020, Lacerna2020, Coccato2022, Maschmann2022, Rathore2022, Raimundo2023, Raimundo2025, Tous2023, Tous2024}. Alternatively, the star formation activity observed in some ETGs could be the aftermath of their recent transition from a late- to an early-type morphology \citep{Martig2009, Sampaio2022}. In contrast, some authors attribute quiescence in LTGs to a previous stage in the morphological transformation into early type \citep[e.g.,][]{Pozzetti2010}, especially when they are observed in high-density environments \citep[e.g.,][]{Kelkar2019, Martinez2023, Oxland2024}. Thus, if we want to build a comprehensive picture of galaxy evolution, we must understand the relationship between morphology and star formation, and be able to explain the more exotic configurations mentioned in this paragraph.

The advent of large galaxy surveys, such as the Sloan Digital Sky Survey \citep[SDSS;][]{York2000} or the Dark Energy Survey \citep{DES}, along with the creation of catalogues of galactic morphologies \citep[e.g.,][]{Nair&Abraham2010, HDS2018, Fischer2019, Vega-Ferrero2021, Vazquez-Mata2022, Walmsley2022, Walmsley2023, Bom2024} has enabled to address these questions with sound statistics. Although these catalogues are useful tools, essential for this endeavour, and have provided robust results, we should be aware of their limitations and not use them blindly. In this letter, we want to present results that illustrate what is, perhaps, the strongest limitation in any visual morphological identification: the apparent inclination of galaxies. We show that, when combined with different morphological classifications, common physically-motivated selection criteria can bias a sample in favour of edge-on galaxies, whose morphological identification is unreliable. Specifically, we report that the selection of quiescent LTGs (Section~\ref{S:sample}) results in a clear excess of edge-on systems (Section~\ref{SS:identification}), which may lead to overestimate the fraction of these galaxies (Section~\ref{SS:impact}). These results lead us to propose some simple measures to mitigate this inclination bias (Section~\ref{SS:mitigation}), and to conclude that the excess of edge-on quiescent LTGs could be attributed to misclassified edge-on ETGs (Section~\ref{S:conclusions}).

Throughout this work, we assume a flat $\Lambda$CDM cosmology, with parameters $H_0 = 70\;{\rm km\;s^{-1}\;Mpc^{-1}}$ and $\Omega_m = 1 - \Omega_\Lambda = 0.3$.

\section{Sample selection}
\label{S:sample}
The sample of galaxies analysed in this work is drawn from the NASA-Sloan Atlas \citep[NSA;][]{Blanton2011} catalogue, which is based on the SDSS main galaxy sample \citep{Strauss2002}. In its last version (v\_1\_0\_1), the NSA catalogue includes virtually all the galaxies with elliptical r-band Petrosian
magnitudes $\leq 17.77$ and known spectroscopic redshifts up to $z = 0.15$ within the coverage of the SDSS data release 13 \citep{Albareti2017}, and incorporates better photometric analysis than the standard SDSS pipeline. From this catalogue, we take the axis ratio ($b/a$) from 2D single-component Sérsic fits in the $r$-band, which we use as a proxy for inclination. To obtain star formation rates (SFRs) and stellar masses ($M_*$), we crossmatch the NSA catalogue with the latest version of the GALEX-SDSS-WISE Legacy Catalogue \citep[GSWLC;][]{Salim2018}, where these parameters were derived by fitting the observed spectral energy distributions of the galaxies across the UV, optical, and IR domains.

To distinguish between LTGs and ETGs, we rely on the morphological catalogue\footnote{The morphological catalogue is available in \url{https://archive.cefca.es/ancillary_data/sdss_morphological_catalogues/sdss_morphological_catalogues.tar.gz}.} of \citet{HDS2023}.  In that work, galaxies were classified from SDSS colour images using the same convolutional neural network model introduced in \citet{HDS2022}, now applied to the full SDSS dataset\footnote{Identical results to the ones presented in this letter are obtained by using the first version of the morphological catalogue \citep{HDS2018}.}. The model was trained on the visual classifications of \citet{Nair&Abraham2010} and the Galaxy Zoo 2 project \citep{Willett2013}. This catalogue provides a set of parameters to characterise the morphology of galaxies, from which we take the probability of being LTG ($P_{\rm LTG}$), the T-type, and the probability of being S0 ($P_{\rm S0}$) which helps to distinguish between elliptical (E) and S0 morphologies. In addition, we also retrieve from this catalogue the probability that a galaxy is seen edge on ($P_{{\rm edge}}$). From the 641,409 galaxies in the NSA catalogue, 451,819 have counterparts in GSWLC, while 494,116 have counterparts in the morphological catalogue. Our main galaxy sample comprises 442,872 objects, corresponding to the intersection of all three catalogues. 

We select LTGs by requiring $P_{\rm LTG} \geq 0.5$, which results in $241,\!321$ objects. To identify quiescent systems, we follow \citet{Tous2024}. We take as the threshold of quiescence an offset of $-1.1$ dex from the main sequence of star-forming galaxies as defined in \citet{Renzini2015}. Therefore, quiescent galaxies must satisfy the following condition:
\begin{equation}
    \log\left[\dfrac{\rm SFR}{\rm M_\odot\; yr^{-1}}\right] - 0.76 \log\left[\dfrac{M_*}{\rm M_\odot}\right] + 7.6 < -1.1\;.
\label{eq:quiescence}
\end{equation}
By imposing these two conditions, we find $14619$ quiescent LTGs in our sample, corresponding to $3$ and $6$ per cent of the main sample and the LTGs, respectively. Note that the conclusions of this work do not change if stricter $P_{\rm LTG}$ values are used to select LTGs, or if different definitions of quiescence are adopted (see Section~\ref{SS:identification}). In addition, we define two comparison samples of ETGs: one with $85,\!554$ S0 galaxies satisfying $P_{\rm LTG} < 0.5$ and T-type $< 0$ and $P_{\rm S0} \geq 0.5$, and another with $57,\!554$ E galaxies satisfying $P_{\rm LTG} < 0.5$ and T-type $< 0$ and $P_{\rm S0} < 0.5$, which correspond, respectively, to $22$ and $11$ per cent of the main sample.

\section{Results and discussion}
\label{S:results}

In the following sections, we assess the distribution of inclinations of our samples to search for any possible bias that may arise from the proposed selection criteria.

\subsection{Identifying the bias}
\label{SS:identification}

Ideally, if a sample is selected based on a property unrelated to the galaxies’ apparent inclination, one would expect the galaxies to be randomly oriented. To test this, we use the observed axis ratio of the galaxies. This parameter combines intrinsic shape and projection effects, but disc galaxies are relatively thin and nearly axisymmetric, so their projected axis ratios primarily trace inclination. Therefore, if the disc galaxies in our samples are randomly oriented, their axis ratios should follow an approximately uniform distribution\footnote{While this is a good approximation for fast-rotator discs, it is not expected to hold for slow-rotator ETGs, whose intrinsically rounder and often triaxial shapes weaken the connection between projected axis ratio and inclination \citep[e.g.,][]{Weijmans2014, Capellari2016}.}.

\begin{figure}
    \centering
	\includegraphics[width=\columnwidth]{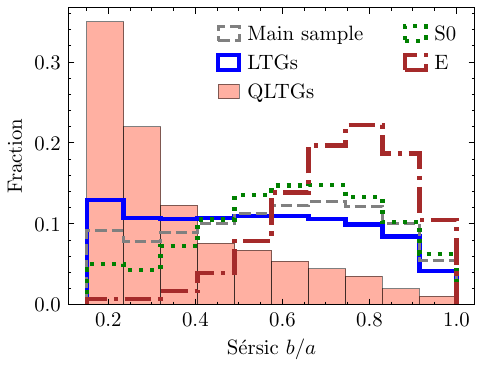}
    \caption{Fraction of galaxies in the different samples as a function of the observed axis ratio. The main galaxy sample is represented by the grey dashed line, late-type galaxies by the blue solid line, quiescent late-type galaxies by the salmon solid boxes, lenticulars by the green dotted line, and ellipticals by the brown dashed-dotted line. The observed axis ratio is used as a proxy for inclination. Compared to the main sample, a clear excess of quiescent late-type galaxies at high inclinations is observed, while S0 galaxies are under-represented.}
    \label{fig:ba}
\end{figure}

Fig.~\ref{fig:ba} shows the fraction of galaxies as a function of the axis ratio for each sample. The main galaxy sample (gray dashed line) includes all morphologies: disc galaxies (LTGs and S0, represented by the solid blue and green dotted lines, respectively) and E galaxies (brown dashed-dotted line), whose distribution is skewed towards higher $b/a$ due to their intrinsically rounder shapes and triaxiality. Nevertheless, the overall distribution is approximately uniform, as it is largely dominated by disc galaxies (Section~\ref{S:sample}). Likewise, the LTG sample shows a roughly flat $b/a$ distribution, as expected for a population of randomly oriented thin discs. However, as shown by the salmon filled histogram, this is not the case for our sample of quiescent LTGs. Their axis ratio distribution is clearly skewed towards low values: it peaks at $b/a \approx 0.2$, whereas around sixty per cent of the galaxies in that sample exhibit axis ratios lower than $0.4$, suggesting that most are highly-inclined discs. This excess of quiescent LTGs with low axis ratios likely explains part of the differences between the distributions of the main galaxy sample and the LTG sample. Indeed, by applying a two-sample Kolmogorov–Smirnov (KS) test to compare both---the main and the LTG---samples, we obtain a low probability, $p << 0.001$, which indicates that it is very unlikely to observe these differences if both samples were drawn from the same parent population. Thus, the sample of quiescent late-type systems likely accounts for the higher fraction of LTGs observed at $b/a < 0.4$ compared to the main galaxy sample. Instead, by setting eq.~\ref{eq:quiescence} to $> -0.5$ to select only star-forming LTGs, we have verified that the $b/a$ distribution looks more similar to that of the main galaxy sample. Residual differences between these distributions can be attributed to the presence of ETGs with high axis ratios in the main sample, which are mostly excluded from the LTG sample.

\begin{figure}
    \centering
	\includegraphics[width=\columnwidth]{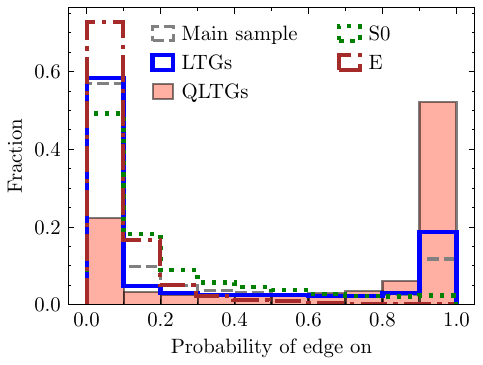}
    \caption{Fraction of galaxies in the different samples as a function of the probability of being edge on \citep{HDS2023}. The main galaxy sample is represented by the grey dashed lined, late-type galaxies by the blue solid line, quiescent late-type galaxies by the salmon solid boxes, lenticulars by the green dotted line, and ellipticals by the brown dashed-dotted line. The fraction of edge-on, quiescent, late-type galaxies is much higher than that of the main galaxy sample, while that fraction for the S0 galaxies is much lower. This supports the idea that the extra edge-on, quiescent, late-type galaxies might be misclassified S0 galaxies.}
    \label{fig:pedge}
\end{figure}

In the catalogue by \citet{HDS2023}, the parameter $P_{\rm edge}$ can be used to identify those systems that are seen edge on. In Fig.~\ref{fig:pedge}, we show the distributions of this parameter for the same galaxy sets: main galaxy sample (gray dashed line), LTGs (blue solid line), quiescent LTGs (red solid bars), S0 galaxies (green dotted line) and E galaxies (brown dashed-dotted line). In line with the conclusions drawn from the axis ratio in Fig.~\ref{fig:ba}, this figure reveals a clear excess of quiescent LTGs with high probability of being viewed edge on, compared to the main galaxy sample or the LTGs. We find that the fraction of quiescent LTGs with high edge-on probability is, indeed, close to sixty per cent.

The large fraction of quiescent LTGs observed at high inclinations (about sixty per cent) suggests that the morphology assigned to many of these systems may be incorrect. An edge-on spiral galaxy is very difficult to distinguish from an edge-on S0 galaxy because it is not possible to determine the presence of spiral arms. This ambiguity is enhanced when the inclined disc is quiescent, since red spiral galaxies are thought to represent a transitional phase between late-type and early-type morphologies, making them similar in appearance and therefore difficult to distinguish based on morphology alone \citep{Bamford2009}. In fact, we find that S0 galaxies are under-represented at $b/a < 0.4$. As indicated by the green dotted histogram in Fig.~\ref{fig:ba}, the fraction of S0 galaxies drops below that of the main galaxy sample at low axis ratios. A KS test indicates that the axis‑ratio distributions of the S0 and main galaxy samples are significantly different (with $p << 0.001$). In principle, any morphologically selected sample of S0 galaxies should contain a substantial fraction of fast-rotators whose intrinsic flattening is around $0.25$ \citep[e.g.;][]{Weijmans2014}. Therefore, the deficit of S0 galaxies with low axis ratio suggests that some might be missing, possibly because they have been confused for quiescent spiral galaxies.

 We reach a similar conclusion by studying the distribution of $P_{\rm edge}$ values for S0 galaxies (green dotted line in Fig.~\ref{fig:pedge}). While the absence of Es (brown dashed-dotted line in Fig.~\ref{fig:pedge}) with high $P_{\rm edge}$ is expected due to their triaxiality and higher axis ratios (Fig.~\ref{fig:ba}), the fraction of inclined S0 galaxies (i.e., early-type discs) should be much higher. In addition, as discussed in Appendix~\ref{A:appendix}, edge-on quiescent LTGs occupy a similar position in the SFR vs $M_*$ diagram as S0 galaxies. Supporting the confusion hypothesis, the T-types assigned by \citet{HDS2023} to quiescent LTGs range from zero to four, coinciding with the range where \citet{HDS2022} report a drop in the number of objects with a reliable visual classification (see figure $8$ in that work). These results are consistent with the findings of \citet{Bom2024}, who identified a fraction of red LTGs with high inclinations in their classification, and suggested that they could be misclassified S0 galaxies.

 It is worth noting that the existence of S0 galaxies that can be easily confused with LTGs when seen edge-on suggests that some may correspond to a subset of lenticular galaxies with structural properties similar to those of spiral galaxies. Given the diversity of evolutionary pathways proposed for S0 galaxies \citep[e.g.,][]{mckelvie2018, Deeley2020, Coccato2022}, such systems could plausibly be associated with disc-dominated S0s formed through the stripping of spiral galaxies in clusters, as discussed, e.g., in \citet{Bom2024}. At the same time, we expect that, in addition to these disc-dominated S0 galaxies, the sample also includes a genuine population of quiescent LTGs with distinct properties \citep{Bamford2009}.

We stress that the inclination bias we report is not simply due to highly inclined star-forming discs being reddened by dust and therefore classified as quiescent. In \citet{Salim2018}, the SFR and $M_*$ were carefully estimated by taking into account the effect of dust attenuation. Besides, we have verified that the inclination bias is still present when quiescent galaxies are selected by using the spectral index $D4000$. This spectral index is measured from the ratio of the continuum flux in two narrow, adjacent wavelength intervals around $4000$ \AA, so it is a proxy for the stellar population age and is largely insensitive to dust attenuation. Thus, we conclude that the observed excess of quiescent LTGs most likely stems from the morphological identification.

\subsection{Impact of the bias}
\label{SS:impact}

Next, we use the axis ratio distribution of the main galaxy sample as an unbiased reference to estimate the fraction of potentially misclassified quiescent LTGs, $\varphi_{\rm mis}$. The observed distribution of quiescent LTGs, $F_{\rm QLTG}(b/a)$, can be expressed as a mixture of the intrinsic LTG distribution represented by the main galaxy sample, $F_{\rm MGS}(b/a)$, and the distribution of misclassified systems, $F_{\rm mis}(b/a)$:
\begin{equation}
    F_{\rm QLTG}(b/a) = (1 - \varphi_{\rm mis})\;F_{\rm MGS}(b/a) + \varphi_{\rm mis}\;F_{\rm mis}(b/a)\;,
    \label{eq:mixture}
\end{equation}
so, solving for $\varphi_{\rm mis}$ gives:
\begin{equation}
    \varphi_{\rm mis} = \dfrac{F_{\rm QLTG}(b/a) - F_{\rm MGS}(b/a)}{F_{\rm mis}(b/a) - F_{\rm MGS}(b/a)}\;.
  \label{eq:fraction_misclassified}
\end{equation}
Assuming that misclassified galaxies populate exclusively the region $b/a < 0.4$, the integrated distribution $F_{\rm mis}$ over this interval equals unity, while $F_{\rm QLTG}$ and $F_{\rm MGS}$ reduce to the observed fractions of galaxies with $b/a < 0.4$ in the quiescent LTG and main galaxy samples, respectively\footnote{These fractions are $ 0.69$ for the quiescent LTGs, and $0.25$ for the main galaxy sample.}. Equation~\ref{eq:mixture} therefore provides a direct estimate of the global fraction of contaminants within the quiescent LTG sample. Accordingly, we find that over $58$ per cent of quiescent LTGs could be misclassified. Nevertheless, the impact of this bias obviously depends on the size of the sample being considered. For example, such a fraction only represents less than four per cent of all the LTGs, and less than two per cent of our main galaxy sample.

Without accounting for this bias, the observed fraction of quiescent LTGs in the main galaxy sample is $\varphi_{obs}\approx 3.3$ per cent. However, if one excludes the two per cent of potentially misclassified systems, the true fraction of quiescent LTGs, estimated as
\begin{equation}
  \varphi_{\rm true} =\left(1 - \varphi_{\rm mis}\right)\,\varphi_{obs}\;,
  \label{eq:fraction_true}
\end{equation}
is much smaller: only about $1.4$ per cent. Consequently, the number of quiescent LTGs selected from morphological catalogues, without accounting for this bias, could be overestimated by a factor of $\approx 2.4$. Likewise, for a volume-limited sample, made of all galaxies at $z < 0.105$ that are brighter than $-20.5$ in the $r-$band absolute magnitude, we find $\varphi_{\rm obs} \approx 3.7$ and $\varphi_{\rm true} \approx 1.8$, implying that the fraction of quiescent LTGs could be overestimated by over a factor of two.

The inclination bias that we have identified seems to be present whenever LTGs are selected based on characteristics that are more typical of early-type morphologies. For instance, we find evidence that this sort of misclassification is also seen when LTGs with kinematically misaligned gas are selected\footnote{Kinematically misaligned galaxies are systems with a large offset between the kinematic position angles of the gas and the stars. Misaligned gas is more frequently observed in ETGs \citep[e.g.][]{Raimundo2023}, where its angular momentum is less efficiently dissipated due to smaller reservoirs of native gas \citep[e.g.][and references therein]{Baker2025}. In S0 galaxies, such kinematic features could be associated with mergers \citep{Deeley2020, Rathore2022}.}, or when, as mentioned earlier in this section, LTGs with old stellar populations are selected. Additionally, we find evidence suggesting that this bias could also be present in other independent morphological classifications. For example, a similar distribution of axis ratios peaking at low values is obtained if one takes the T-types from \citet{Vazquez-Mata2022} to select LTGs. Therefore, beyond standard recommendations to treat the morphology of edge‑on galaxies with caution \citep[as noted, e.g., in section 4 of][]{Vega-Ferrero2021}, our results further suggest that the inclination distribution should be explicitly evaluated for any LTG sample selected on the basis of early‑type–like properties.

\subsection{Guidelines to mitigate the inclination bias}
\label{SS:mitigation}

Fortunately, many catalogues provide useful tools that can be used to identify and mitigate this inclination bias. Among these tools, we note that the most reliable ones are those parameters derived from objective measurements---such as the observed axis ratio, or edge-on identifications---rather than quality flags that come from visual inspection. For instance, we have shown that $P_{\rm edge}$ from \citet{HDS2023} reveals the excess of highly-inclined, quiescent LTGs. Likewise, \citet{Vazquez-Mata2022} flagged as edge on around forty per cent of quiescent LTGs in MaNGA. In contrast, the vast majority of quiescent LTGs in MaNGA are flagged as visually confirmed spirals with reliable morphology by \citet{HDS2022} or \citeauthor{Vazquez-Mata2022}. The latter authors have used a visual estimate of the bulge-to-total light ratio to distinguish between edge-on spiral and S0 galaxies. However, this test is not bulletproof because the light profile of both, spiral and S0 galaxies, can be quite similar, as pseudo bulges are relatively common among both Hubble types \citep[e.g.,][]{Laurikainen2007, Tous2020, HDS2022}. 

Therefore, the simplest and most conservative solution to remove this bias from any sample is to exclude edge-on galaxies. Alternatively, one could derive a statistical correction to weight each quiescent LTG as a function of its inclination to make its probability density function match that of an unbiased reference sample. In our case, we could fit a function $f(b/a)$ to the frequency of quiescent LTGs in Fig.~\ref{fig:ba} and another function $g(b/a)$ to the main galaxy sample, which would be the reference, and then derive the weight:
\begin{equation}
    w(b/a) = \dfrac{g(b/a)}{f(b/a)}\;.
\end{equation}
Nevertheless, since the purpose of this letter is to raise awareness about this bias---and such a correction is sample dependent---we refrain from providing explicit expressions for these functions.

\section{Conclusions}
\label{S:conclusions}

In this work, we present evidence that commonly used selection criteria for identifying quiescent LTGs may result in a significantly biased sample, in which over $58$ per cent of the galaxies may instead be misclassified early‑type discs (i.e., S0) seen edge‑on. This bias becomes apparent when examining the observed axis ratio distribution of quiescent LTGs, which shows a pronounced excess of systems at low values compared to both the parent population of all LTGs and the unbiased NSA catalogue that includes all morphologies. In contrast, S0 galaxies---early‑type discs lacking spiral arms---appear under‑represented at high inclinations, suggesting that a fraction of them may have been confused with quiescent LTGs, hence contributing to the observed excess. The similar location of edge‑on quiescent LTGs and S0 galaxies in the star formation rate–stellar mass plane further supports this conclusion. 

We find evidence suggesting that this bias may be present in different morphological catalogues, and that it is unrelated to the specific quiescence criteria used. Our analysis suggests that the true fraction of quiescent LTGs may be overestimated by more than a factor of two if inclination effects are not taken into account. To mitigate this bias, we recommend either removing edge‑on galaxies from the sample, or implementing a statistical correction using the axis‑ratio distribution of an unbiased reference data set---such as the NSA catalogue when working with SDSS galaxies (see Section~\ref{SS:mitigation}).

We conclude that the bias here reported likely originates from the inherent difficulty of reliably determining morphology in highly inclined systems: without a prominent classical bulge, edge‑on discs are easily misclassified as late‑type galaxies. Since the neural network used by \citet{HDS2023} was trained on human visual classifications, our findings reflect a broader limitation of many artificial intelligence approaches: that models inevitably inherit the biases present in their training data sets. Finally, we emphasize that this bias may remain unnoticed unless the distribution of inclinations of a sample is explicitly examined. We therefore encourage future studies to follow our recommendations when selecting quiescent LTGs.

\section*{Acknowledgements}
The authors thank the anonymous referee for a constructive report that helped improve the presentation of our results. JL would like to extend his thanks to Jesús Vega Ferrero for useful discussions on the morphological classifications. This work was supported by the Science and Technology Facilities Council (STFC) of the UK Research and Innovation via grant reference ST/Y002644/1. HDS acknowledges financial support by RyC2022-030469-I grant, funded by MCIU/AEI/10.13039/501100011033.

\section*{Data availability}
The three public catalogues used in this work are available in: \url{https://nsatlas.org/} (NSA),  \url{https://salims.pages.iu.edu/gswlc/} (GSWLC), and \url{https://archive.cefca.es/ancillary_data/sdss_morphological_catalogues/sdss_morphological_catalogues.tar.gz} (morphologies).



\bibliographystyle{mnras}
\bibliography{biblio}



\appendix

\section{Edge-on quiescent LTG\lowercase{s} in the SFR-$\bf{M_*}$ plane}
\label{A:appendix}

Here we study the distribution of edge-on quiescent LTGs (i.e., with $P_{\rm edge} > 0.7$, as given in \citealt{HDS2023}) within the SFR vs $M_*$ plane. 

\begin{figure}
    \centering
	\includegraphics[width=\columnwidth]{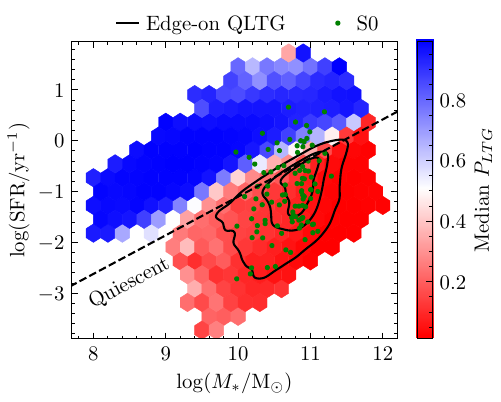}
    \caption{Star formation rate as a function of stellar mass for edge-on quiescent late-type galaxies (black contour lines) and S0 galaxies (green dots). Contours are derived from a kernel density estimate of the distribution of edge-on quiescent late-type galaxies. To avoid overcrowding, only $100$ random S0 galaxies are shown. The background cells show the median probability of being late-type galaxy \citep{HDS2023}, inferred from the main galaxy sample, with blue meaning higher probability and red lower probability. The black dashed line marks the quiescent threshold (eq.~\ref{eq:quiescence}). The apparent contour extension above this threshold is an artefact of the kernel density estimate, as all edge-on quiescent late-type galaxies lie below it. Edge-on quiescent late-type galaxies occupy a region dominated by early-type galaxies, which is also the preferred location of most S0.}
    \label{fig:mass_sfr}
\end{figure}

In Fig.~\ref{fig:mass_sfr}, we represent this distribution with a kernel density estimate (black contours) over a background coloured by the median probability of being LTG inferred from our main galaxy sample. This exercise reveals that edge-on, quiescent LTGs are located at rather high $M_*$ ($\approx 10^{10.8}\;{\rm M_\odot}$), close to the quiescence threshold (eq.~\ref{eq:quiescence}), in a zone dominated by ETGs. This location coincides quite well with the one preferred by most S0 galaxies, which are represented by green dots in the figure. This coincidence further supports the idea that the excess of quiescent LTGs at high inclinations revealed by Fig.~\ref{fig:ba} could be explained if most of these systems are misclassified edge-on S0 galaxies.

\bsp	
\label{lastpage}
\end{document}